\documentclass[lettersize,journal]{IEEEtran}
\usepackage{blindtext,cite,hyperref}
\usepackage{ctable,amsmath,amssymb,amsfonts}
\usepackage{algorithmic}
\usepackage{algorithm}
\usepackage{graphicx}
\usepackage{textcomp}
\usepackage{makecell}
\usepackage{float}
\usepackage{placeins}
\usepackage{caption}
\usepackage{pifont}
\newcommand{\cmark}{\ding{51}}%
\newcommand{\xmark}{\ding{55}}%
\begin{document}
\title{Inverse Problem of Ultrasound Beamforming with Denoising-Based Regularized Solutions}
\author{Sobhan Goudarzi, Adrian Basarab, \IEEEmembership{Senior Member, IEEE}, Hassan Rivaz, \IEEEmembership{Senior Member, IEEE}
\thanks{This paper has been accepted for publication in IEEE Transactions on Ultrasonics, Ferroelectric, and Frequency Control.}
\thanks{This work was supported by Natural Sciences and Engineering Research Council of Canada (NSERC) RGPIN-2020-04612.}
\thanks{Sobhan Goudarzi and Hassan Rivaz are with the Department of Electrical and Computer Engineering, Concordia University, Montreal, QC, H3G 1M8, Canada.\newline
Adrian Basarab is with the Universit\'e de Lyon, INSA-Lyon, UCBL, CNRS, Inserm, CREATIS UMR 5220, U1206, Villeurbanne, France.\newline
Email: sobhan.goudarzi@concordia.ca~, adrian.basarab@irit.fr~, hrivaz@ece.concordia.ca~}}
\maketitle
\begin{abstract}
During the past few years, inverse problem formulations of ultrasound beamforming have attracted a growing interest. They usually pose beamforming as a minimization problem of a fidelity term resulting from the measurement model plus a regularization term that enforces a certain class on the resulting image. Herein, we take advantages of alternating direction method of multipliers to propose a flexible framework in which each term is optimized separately. Furthermore, the proposed beamforming formulation is extended to replace the regularization term by a denoising algorithm, based on the recent approaches called plug-and-play (PnP) and regularization by denoising (RED). Such regularizations are shown in this work to better preserve speckle texture, an important feature in ultrasound imaging, than sparsity-based approaches previously proposed in the literature. The efficiency of proposed methods is evaluated on simulations, real phantoms, and \textit{in vivo} data available from a plane-wave imaging challenge in medical ultrasound. Furthermore, a comprehensive comparison with existing ultrasound beamforming methods is also provided. These results show that the RED algorithm gives the best image quality in terms of contrast index while preserving the speckle statistics.  
\end{abstract}
\section{Introduction}
\label{sec:sec1}
\IEEEPARstart{M}{edical} ultrasound probe is made of several piezoelectric elements used for the transmission of non-invasive acoustic waves into the medium and also the reception of the backscattered signals. To insonify the medium with a desired wave, the excitation pulses of transducer elements as well as their firing times are adjusted in a process called transmit beamforming~\cite{cobbold2006}. The backscattered echoes from each location of the medium are traced back in receive beamforming to reconstruct a spatial map of the tissue echogenicity~\cite{cobbold2006}.\par
Although receive beamforming is an ill-posed inverse problem, delay-and-sum (DAS) algorithm is commonly used for real-time ultrasound imaging in commercial scanners. DAS simply provides the backprojection solution of the inverse problem of beamforming~\cite{7081460}. Nevertheless, DAS uses a predefined apodization window for the entire image which limits the resulting quality by the well-known trade-off between the level of side lobes and the width of the main lobe in the frequency domain. Adaptive beamforming methods have been developed to effectively determine the	apodization weights based on the echo signals~\cite{5278437,849225,6960091}.\par
Inverse problem formulation of ultrasound beamforming is another alternative to DAS, wherein a measurement model is considered for the synthesis of the desired image~\cite{6203608,8636261,8091286,7565515,8052532}. Linear models relate each sample of received channel data to pixels of the image to be recovered through a weighting matrix. While linear models are simple and provide a plausible approximation of the image under scrutiny, the weighting matrix is usually of size several hundreds of thousands which makes the model memory intensive. Despite its high dimension, the weighting matrix is usually sparse and thus easy to store. However, the high dimension of the inverse problem to be solved requires optimization algorithms and representations (e.g., operators) that do not include matrix inversion~\cite{7565515,8052532,8091286,2021unifying}.\par
Similar to most inverse problems in computational imaging, the existing inverse problem-based beamformers in ultrasound imaging use regularization functions derived from \textit{a priori} statistics of the ultrasound image. The most used are Gaussian models, turning into an $\ell_2$-norm regularization term, or Laplacian promoting sparsity through the $\ell_1$-norm~\cite{7565515,8052532,8091286,2021unifying,7174535}. While the latter is well-adapted to reconstruct ultrasound images with high resolution and contrast, sparse solutions have shown a poor performance in preserving the speckle texture, which is an important feature for applications such as motion estimation and tissue classification.\par
As an alternative to standard regularization functions used to solve image reconstruction or restoration problems, an important class of methods has been proposed in the computational imaging literature~\cite{7744574,6737048,7542195,16M1102884}. The main idea is to use denoising algorithms as regularizers. Specifically, Venkatakrishnan~\textit{et al.} proposed an interesting idea termed plug-and-play (PnP)~\cite{6737048} based on the alternating direction method of multipliers (ADMM) which allows decoupling the measurement model and the regularization terms. It has been shown that the proximity operation related to the regularization term can be replaced by an image denoising algorithm~\cite{6737048}. This idea has been increasingly applied to a number of applications~\cite{7744574,7542195,9454311,8434327}. Nevertheless, the explicit objective function of PnP approach is unknown, and this issue strongly limits studying theoretical convergence properties and interpretations. New insights on characterization of proximity operator and the proof of convergence for the PnP algorithm under certain conditions have been presented in~\cite{hurault2021,97108695,hurault2022}.\par
Recently, Romano~\textit{et al.} proposed a skillful way to plug-in denoising algorithms when solving imaging inverse problems called regularization by denoising (RED)~\cite{16M1102884}. The explicit regularizer of RED is designed to enforce the orthogonality of the image and what a denoiser removes from the image. It has been shown that for locally homogeneous denoisers, the gradient expression of regularizer can be easily found, and several iterative algorithms were proposed to find the optimal solution~\cite{16M1102884}. Reehorst and Schniter~\cite{8528509} shed more light on the RED algorithm and have shown that besides local homogeneity, the denoising algorithm must also be Jacobian symmetric in order to be explained by an explicit regularization term. Although RED works very well in practice, it has been shown that common denoisers lack Jacobian symmetry property. Therefore, in~\cite{8528509} RED has been explained in a novel framework called score-matching by denoising (SMD).\par 
Herein, inspired by the success of PnP and RED algorithms in various medical imaging inverse problems~\cite{100228756,8434327,18M1169655,8353870}, we devise a general framework for the inverse problem of ultrasound beamforming based on the ADMM. We use a linear forward model for the image under scrutiny, and the basic solution is found by considering $\ell_1$-norm regularizer which we will refer to henceforth as ADMM solution. Moreover, the proposed framework is extended with both PnP and RED algorithms refereed to henceforth as PnP and RED solutions, respectively. The performance evaluation is presented on simulations, real phantoms, and \textit{in vivo} datasets available from the plane-wave imaging challenge in medical ultrasound (PICMUS)~\cite{7728908}. Furthermore, a comprehensive comparison with other beamforming approaches is also provided. Our main contributions can be summarized as follows:
\begin{enumerate}
	\item Inverse problem of ultrasound beamforming is formulated in a flexible framework based on ADMM.
	\item For the first time, denoising-based regularization functions are applied to ultrasound beamforming problem.
	\item The solution of measurement model is obtained using numerical methods, which does not involve the intractable calculation of Hessian matrix.
	\item Comprehensive experiments on different datasets are performed to evaluate the performance of proposed method in terms of quantitative metrics, and also in comparison with other approaches.
\end{enumerate} 
Furthermore, the source codes for Matlab implementations of the proposed algorithms are publicly available in these links:~\url{https://github.com/Sobhan-Goudarzi/Denoising-Based-Ultrasound-Beamforming} and~\url{code.sonography.ai}.
\subsection{Related work}
\label{sec:sec11}
During the past few years, inverse problem formulations have attracted a growing interest in the field of medical ultrasound imaging. They have been used in different problems such as deconvolution~\cite{7302609,9210758}, despeckling~\cite{1588392,5428848}, compressive sensing (CS)~\cite{LIEBGOTT2013525,7118193,4919302,7532811,6863846}, and beamforming~\cite{7565515,8052532,8091286}. In particular, CS applications perhaps popularized the inverse problem-based approaches in ultrasound imaging. The CS reconstruction of radio frequency (RF) channel data was performed in~\cite{LIEBGOTT2013525}. Afterward, CS was applied on beamformed envelope data to reduce the size of stored data~\cite{7118193}. CS has also been used to reconstruct a high-quality ultrasound image by using a reduced number of transducer elements~\cite{4919302,7532811} or sub-Nyquist sampled data~\cite{6863846}. Later, beamforming has been formulated as an inverse problem in order to improve the quality of ultrasound images. Szasz~\textit{et al.} assumed a linear model between the RF channel data and the desired image~\cite{7565515}. Each image depth was then reconstructed by solving a regularized inverse problem, with applications to both plane-wave~\cite{7728907} and focused imaging~\cite{7565515}. This approach was further extended by adding more regularization terms and reconstructing all image depths concurrently~\cite{8052532}. Besson~\textit{et al.} developed two matrix-free formulations to mitigate the memory and computational requirements of the inverse problem of ultrafast imaging~\cite{8091286}.\par
Recently, deep learning has become another option for solving the ultrasound beamforming problem~\cite{101121,8302520,8663450,9314245,9178454}. Deep models make ill-posed inverse problems tractable thanks to their great potential for approximating non-linear mapping functions between high dimensional training pairs. Convolutional neural networks (CNNs) were adapted to estimate high quality In-phase/Quadrature (IQ) data from delayed RF data~\cite{9025198}. Generative adversarial networks (GANs) were used to mimic eigenspace-based minimum variance (EMV) beamforming~\cite{ZHOU2021102086}. The challenge on ultrasound beamforming with deep learning (CUBDL) was held in conjunction with the 2020 IEEE International Ultrasonics Symposium (IUS)~\cite{9475029}. In~\cite{9251565}, deep learning was adapted to construct a general ultrasound beamformer and successfully taken apart in this challenge. This method was designed to mimic minimum variance (MV) beamforming using MobileNetV2 network architecture, and was ranked first in terms of image quality. Overall, considering the network size as well, it was jointly ranked first with another submission~\cite{9251322}. Recently, self-supervised learning~\cite{ZHANG2021102018} as well as Complex CNNs~\cite{9614147} were adapted for plane-wave beamforming. Deep learning has also been used to reconstruct ultrasound images from sub-sampled data~\cite{9738614,9467278}.\par
\section{Inverse problem of ultrasound beamforming}
\label{sec:sec2}
\begin{figure}[t]
	\captionsetup{justification=centering}
	\centerline{\includegraphics[width=8.5cm]{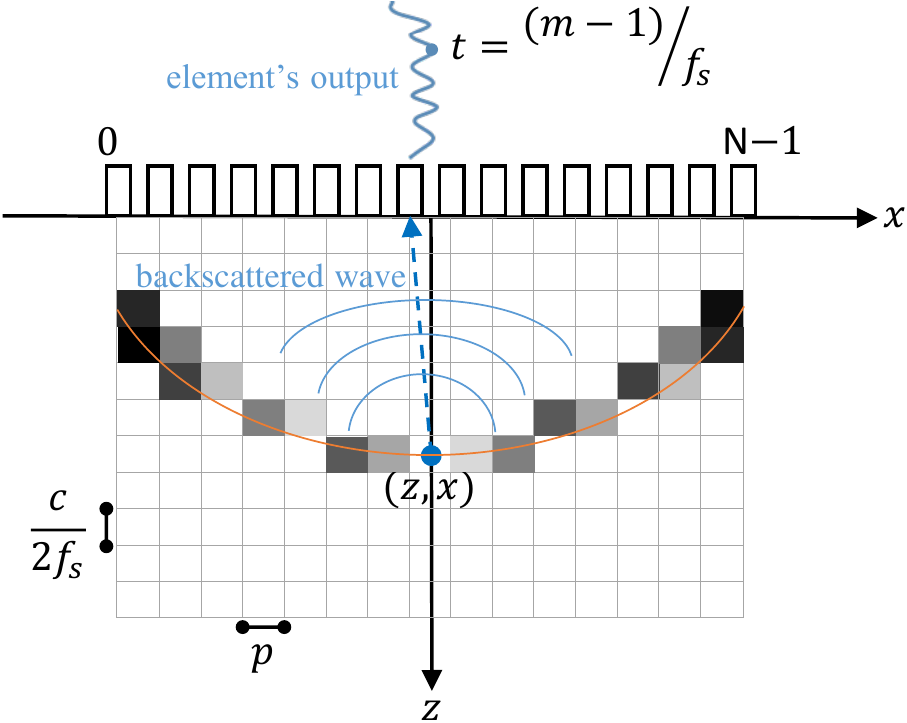}}
	\caption{The illustration of image pixels' contribution into a single sample of pre-beamformed data.}
	\label{fig:fig4}
\end{figure}

The goal of ultrasound imaging is to form a high-quality spatial map of the medium echogenicity. To do so, $L$ piezoelectric elements of an ultrasound probe transmit an acoustic wave into the medium, and the backscattered waves are collected with $N$ crystal elements of the same probe. Depending on the imaging technique (e.g., line-per-line, plane-wave, or synthetic aperture imaging) and the probe type (e.g., linear, curvilinear, or phased array), this process may be repeated several times to form a single image. Fig.~\ref{fig:fig4} shows an example for N-element linear probe, with a transducer pitch of $p$, in which the backscattered signals are recorded with a specific sampling frequency ($f_s$), and the beamforming grid is broken into certain numbers of pixels in the axial (i.e., the wave propagation direction) and lateral directions $(z,x)$ with the pixel sizes of $d_z = \frac{c}{2f_s}$ and $d_x = p$, respectively. The speed of sound in the medium is assumed constant and denoted by $c$. If the time offset following a transmit event equals zero, the actual time corresponding to $m^{th}$ sample of recorded signals is $t=(m-1)/f_s$, where $m=\{1, 2, ..., M\}$.\par
During the reception, echoes from different pixels might simultaneously arrive at a transducer element and lead to a single output sample if and only if the sum of propagation times for transmitted wave reaching them ($\tau_t$) and getting back to that element ($\tau_r$) was the same. Considering digitization error, all pixels respecting the following condition contribute to that sample of element's output:
\begin{equation} 
\label{eq:2}
\mid t-\tau \mid \leq \frac{1}{f_s} ,
\end{equation}
where $\tau=\tau_t+\tau_r$ is the propagation delay of each pixel and depends on its location, the probe geometry, the type of transmitted ultrasound wave (e.g., focused, plane-wave, or spherical wave), and assumed speed of sound. As illustrated in Fig.~\ref{fig:fig4}, Eq.~(\ref{eq:2}) results in an elliptical region with varying weights. 
Therefore, each sample of the RF channel data can be linearly modeled as a combination of the pixels' values in the desired image. The forward model can be written as follows: 
\begin{equation} 
\label{eq:3}
\mathbf{y} = \Phi \mathbf{x} + \mathbf{\nu} ,
\end{equation}
where $\mathbf{y}, \mathbf{x}\in\mathbb{R}^{MN}$ are the vectorized versions of pre-beamformed data and desired image, respectively. $\Phi\in\mathbb{R}^{MN\times MN}$ stands for the weighting matrix, and $\mathbf{\nu}$ is the electronic noise affecting the raw data which has been shown to be well-approximated by additive white Gaussian noise (AWGN)~\cite{wagner1987}.\par
The way, used in here, for designing matrix $\Phi$ has been inspired by the work in~\cite{8052532}. This simple method is based on the assumption of linear propagation in the medium and does not incorporate the Point Spread Function (PSF) of the probe.
As shown in Fig.~\ref{fig:fig4}, each row of matrix $\Phi$ includes the contributions of image pixels into a single sample of pre-beamformed data. More specifically, the pixels are weighted using the following equation:
\begin{equation} 
\label{eq:1}
\Phi(i,j) = \begin{cases}1-\frac{\mid t_i-\tau_j \mid}{t_{max}} & \mid t_i-\tau_j \mid \leq \frac{1}{f_s}\\0 & \mid t_i-\tau_j \mid > \frac{1}{f_s}\end{cases},
\end{equation}
where $t_{max}$ is the maximum absolute difference between the actual time ($t_i$) corresponding to a sample of element's output and the propagation delays ($\tau_j$) of pixels which contribute to that sample (i.e., only the ones respecting the $\mid t_i-\tau_j \mid \leq \frac{1}{f_s}$ condition). Matrix $\Phi$ is highly sparse because only a small portion of pixels satisfy Eq.~(\ref{eq:2}). Moreover, $\Phi$ is data independent and can be precalculated based on the known imaging settings. Finally, matrix $\Phi$ is multiplied with a reception apodization matrix, commonly used in DAS beamforming, in which the directionality of transducer elements is taken into account and the f-number is fixed for the entire image. It has to be emphasized that matrix $\Phi$ does not necessarily need to be a square matrix because it can be defined for any grid partitioning not equal to pre-beamformed data. Further explanations regarding the construction of matrix $\Phi$ can be found in the supplementary material.\par
A popular approach, adopted here, to invert Eq.~(\ref{eq:3}) and recover the beamformed image $\mathbf{x}$ is to use a variational approach~\cite{scherzer2008}, wherein we pose and solve the following optimization problem:
\begin{equation} 
\label{eq:4}
\hat{\mathbf{x}} = \operatorname*{argmin}_{\mathbf{x}} \frac{1}{2}\parallel \mathbf{y} - \Phi \mathbf{x} \parallel_2^2+\mu\parallel\mathbf{x}\parallel_1,
\end{equation} 
where the first term is the data loss that penalizes mismatch to the observed RF channel data, and the second term is the regularization term that promotes the sparsity of beamformed image $\mathbf{x}$. Another way of thinking about this optimization problem is to use a Bayesian prospective in which Eq.~(\ref{eq:4}) is equivalent to maximum \textit{a posteriori} estimation of $\mathbf{x}$ with a zero-mean Laplacian prior. Interested readers about the probabilistic interpretation of regularization are referred to~\cite{james2021}. The constant hyperparameter $\mu$ controls the contribution of the data fidelity and sparse regularization terms. Note that the choice of the $\ell_1$-norm to promote sparsity has been extensively used in ultrasound image reconstruction~\cite{7565515,7728907,8052532}. However, other regularization have been also used, such as wavelet frames~\cite{8091286}, $\ell_2$-norm~\cite{7565515}, $\ell_p$-norms~\cite{7174535}, etc. For simplicity, we focus here on $\ell_1$-norm and extend it to newly proposed priors.\par
The objective function presented in Eq.~(\ref{eq:4}) is convex but the $\ell_1$ term is nondifferentiable and the problem does not have a closed-form solution. First-order proximal splitting algorithms~\cite{combettes2011} that operate individually on each term are well suited for this optimization. Herein, ADMM is adopted to find the solution of Eq.~(\ref{eq:4}), whose convergence has been proven for convex optimization problems~\cite{2200000016}.\par
In the reminder of this section, details on solving Eq.~(\ref{eq:4}) using ADMM are first outlined. Afterward, beamforming using PnP algorithm is introduced in Section~\ref{sec:sec22}. Finally, Section~\ref{sec:sec23} describes how Eq.~(\ref{eq:4}) is modified and solved based on the RED algorithm.
\subsection{ADMM solution}
\label{sec:sec21}
Split-variable ADMM is based on optimizing each term of the objective function separately, which is very useful in practice when a single optimization approach cannot be used for all terms~\cite{2200000016}. To do so, the independent variable $\mathbf{x}$ is splitted into two variables $\mathbf{u}$ and $\mathbf{v}$ with the constraint that $\mathbf{u}=\mathbf{v}$. Consequently, the new, but fully equivalent, form of Eq.~(\ref{eq:4}) is as follows:
\begin{equation} 
\label{eq:5}
(\hat{\mathbf{u}},\hat{\mathbf{v}}) = \operatorname*{argmin}_{(\mathbf{u},\mathbf{v})} \frac{1}{2} \parallel \mathbf{y} - \Phi \mathbf{u} \parallel_2^2+\mu\parallel\mathbf{v}\parallel_1 \; s.t.\; \mathbf{u} = \mathbf{v},
\end{equation} 
where the corresponding unconstrained problem can be written using the augmented Lagrangian approach as following:
\begin{multline}
\label{eq:6}
(\hat{\mathbf{u}},\hat{\mathbf{v}},\hat{\mathbf{\lambda}}) = \operatorname*{argmin}_{(\mathbf{u},\mathbf{v},\mathbf{\lambda})} \frac{1}{2} \parallel \mathbf{y} - \Phi \mathbf{u} \parallel_2^2+\mu\parallel\mathbf{v}\parallel_1\\-\mathbf{\lambda}^T(\mathbf{u}-\mathbf{v})+\frac{\beta }{2}\parallel \mathbf{u}-\mathbf{v} \parallel_2^2,
\end{multline}
where $\mathbf{\lambda}\in\mathbb{R}^{MN}$ is the Lagrange multiplier, and $\beta>0$ is the weight of penalty term which penalizes violation from the constraint.\par
The equivalent but more compact form of Eq.~(\ref{eq:6}) is as follows~\cite{Bouman2013}:
\begin{multline}
\label{eq:7}
(\hat{\mathbf{u}},\hat{\mathbf{v}},\hat{\mathbf{\lambda}}) = \operatorname*{argmin}_{(\mathbf{u},\mathbf{v},\mathbf{\lambda})} \frac{1}{2} \parallel \mathbf{y} - \Phi \mathbf{u} \parallel_2^2+\mu\parallel\mathbf{v}\parallel_1\\+\frac{\beta }{2}\parallel \mathbf{u}-\mathbf{v}+\frac{\mathbf{\lambda}}{\beta} \parallel_2^2.
\end{multline}
As mentioned before, ADMM separates the minimization of each term in Eq.~(\ref{eq:7}), and finds its solution through an iterative process as summarized in Algorithm~\ref{alg:1}.
\begin{algorithm}[t]
	\caption{Ultrasound beamforming using ADMM}	
	\begin{algorithmic}[1]
		\label{alg:1}
		\STATE \textbf{Input:}\: $\Phi$, $\mathbf{y}$
		\STATE \textbf{Set:\:} $\mu>0$, $\beta>0$, $\mathbf{u}^{0}$, $\mathbf{v}^{0}$, $\mathbf{\lambda}^{0}$, $\epsilon$
		\STATE \textbf{While\:\:}stopping criterion\:$>\epsilon$\:\:\textbf{do}
		\STATE $\mathbf{u}^{i+1}=\operatorname{argmin}_{\mathbf{u}}\frac{1}{2} \parallel \mathbf{y} - \Phi \mathbf{u} \parallel_2^2+\frac{\beta }{2}\parallel \mathbf{u}-\mathbf{v}^i+\frac{\mathbf{\lambda}^i}{\beta} \parallel_2^2$
		\STATE $\mathbf{v}^{i+1}=\operatorname{argmin}_{\mathbf{v}}\mu\parallel\mathbf{v}\parallel_1+\frac{\beta }{2}\parallel \mathbf{u}^{i+1}-\mathbf{v}+\frac{\mathbf{\lambda}^i}{\beta} \parallel_2^2$
		\STATE $\mathbf{\lambda}^{i+1}= \mathbf{\lambda}^i+\beta(\mathbf{u}^{i+1}-\mathbf{v}^{i+1})$
		\STATE \textbf{End}
	\end{algorithmic}
\end{algorithm}
After setting the hyperparameters $\mu$ and $\beta$, the iterative algorithm is initialized with arbitrary values for the new variables (i.e., $\mathbf{u}$ and $\mathbf{v}$) as well as the Lagrange multiplier ($\lambda$). In each iteration, the cost function presented in Eq.~(\ref{eq:7}) is calculated, and once its relative error for two consecutive iterations becomes smaller than a small constant threshold $\epsilon$, the algorithm ends. The proposed ADMM solution includes two main steps as follows.
\subsubsection{Beamforming update}
\label{sec:sec211}
The first step corresponds to the minimization of least-squares term written in line 4 of Algorithm~\ref{alg:1}. The optimal solution of this cost function can be easily found by setting the gradient to zero, given that it is an unconstrained and differentiable convex problem. By doing so, the following closed-form solution is obtained:
\begin{equation}
\label{eq:8}
\mathbf{u}^{i+1}=({\Phi^T\Phi+\beta J})^{-1}(\Phi^T\mathbf{y}+\beta\mathbf{v}^i-\mathbf{\lambda}^i) ,
\end{equation}
where $J$ is a matrix of ones with the same size as $\Phi^T\Phi$. This solution, however, involves large matrix inversion which is intractable in practice because $\Phi^T\Phi$ is a square matrix of size several hundreds of thousands. $\Phi^T\Phi$ is not even a diagonal matrix, nor one that can be diagonalizable through Fourier transform. Therefore, numerical methods are used to solve this issue and find the solution of this step. Herein, the limited-memory BFGS solver\footnote{MATLAB implementation is publicly available in this link:~\url{http://www.cs.ubc.ca/~schmidtm/Software/minFunc.html}} is used to find the optimal solution. Limited-memory BFGS is a highly efficient quasi-Newton method for unconstrained optimization of differentiable real-valued high-dimensional functions that achieves quadratic convergence for many problems~\cite{Nocedal1}.
\subsubsection{Sparsity and Lagrange multiplier updates}
\label{sec:sec212}
The line 5 of Algorithm~\ref{alg:1} corresponds to the optimization of sparsity constraint. It is commonly referred to as proximal mapping of the $\ell_1$-norm as following~\cite{Bouman2013}:
\begin{multline}
\label{eq:9}
prox_{\mu\parallel.\parallel_1/\beta}(\mathbf{u}^{i+1}+\frac{\mathbf{\lambda}^i}{\beta})=\operatorname{argmin}_{\mathbf{v}}\mu\parallel\mathbf{v}\parallel_1+\\\frac{\beta }{2}\parallel \mathbf{u}^{i+1}-\mathbf{v}+\frac{\mathbf{\lambda}^i}{\beta} \parallel_2^2
\end{multline}
where the objective function is strictly convex, and its optimum solution can be found using shrinkage function~\cite{Bouman2013}, which operates as the soft-thresholding operator:
\begin{multline}
\label{eq:10}
\mathbf{v}^{i+1}=soft_{\mu/\beta}(\mathbf{u}^{i+1}+\frac{\mathbf{\lambda}^i}{\beta})=\\=max\{|\mathbf{u}^{i+1}+\frac{\mathbf{\lambda}^i}{\beta}|-\frac{\mu}{\beta},0\}sign(\mathbf{u}^{i+1}+\frac{\mathbf{\lambda}^i}{\beta}) ,
\end{multline}
Finally, the 6th line of Algorithm~\ref{alg:1} entails updating the Lagrangian multiplier.\par
\subsection{PnP solution}
\label{sec:sec22}
Our primary motivation for using ADMM is to construct a flexible framework which can be extended to PnP and RED algorithms. By closely looking at step 5 of Algorithm~\ref{alg:1}, it can be associated to a variational denoising problem wherein $(\mathbf{u}^{i+1}+\frac{\mathbf{\lambda}^i}{\beta})$ is the observation and $\mathbf{v}$ is the clean image to be estimated. Thanks to the modular structure of the ADMM, step 5 can be accomplished not only using the soft-thresholding function explained before, but any other approach which removes the noise term of the observation is also applicable. This point is the core idea of PnP approach~\cite{6737048} in which the choice of regularizer ($\ell_1$-norm in Eq.~(\ref{eq:4})) is replaced by the choice of any sophisticated denoising algorithm.\par
Motivated by the success and wide applications of nonlocal means (NLM) denoiser in ultrasound imaging~\cite{4982678}, the shrinkage function is replaced herein with NLM algorithm, which is grounded in two reasons. First, NLM has a low number of hyperparameters which are easy to tune. Second, the noise type of ultrasound RF data is additive Gaussian mainly brought about by the sensor noise and the acquisition card~\cite{wagner1987}. Therefore, the Gaussian-weighted Euclidean distance can be reliably used within NLM algorithm to assess the similarity between image patches.\par
\begin{algorithm}[t]
	\caption{Ultrasound beamforming using PnP}	
	\begin{algorithmic}[1]
		\label{alg:2}
		\STATE \textbf{Input:}\: $\Phi$, $\mathbf{y}$
		\STATE \textbf{Set:\:} $\beta>0$, $\mathbf{u}^{0}$, $\mathbf{v}^{0}$, $\mathbf{\lambda}^{0}$, $\epsilon$
		\STATE \textbf{While\:\:}stopping criterion\:$>\epsilon$\:\:\textbf{do}
		\STATE $\mathbf{u}^{i+1}=\operatorname{argmin}_{\mathbf{u}}\frac{1}{2} \parallel \mathbf{y} - \Phi \mathbf{u} \parallel_2^2+\frac{\beta }{2}\parallel \mathbf{u}-\mathbf{v}^i+\frac{\mathbf{\lambda}^i}{\beta} \parallel_2^2$
		\STATE $\mathbf{v}^{i+1}=\mathcal{F}(\mathbf{u}^{i+1}+\frac{\mathbf{\lambda}^i}{\beta})$
		\STATE $\mathbf{\lambda}^{i+1}= \mathbf{\lambda}^i+\beta(\mathbf{u}^{i+1}-\mathbf{v}^{i+1})$
		\STATE \textbf{End}
	\end{algorithmic}
\end{algorithm}
Our PnP beamforming approach is summarized in Algorithm~\ref{alg:2}, wherein the NLM denoiser, denoted by $\mathcal{F}$, is applied to the observation ($\mathbf{u}^{i+1}+\frac{\mathbf{\lambda}^i}{\beta}$) in each iteration. Except for line 5, the rest of the algorithm is exactly the same as ADMM while the hyperparameter $\mu$ does not exist anymore. 
\subsection{RED solution}
\label{sec:sec23}
\begin{algorithm}[b]
	\caption{Ultrasound beamforming using RED}	
	\begin{algorithmic}[1]
		\label{alg:3}
		\STATE \textbf{Input:}\: $\Phi$, $\mathbf{y}$
		\STATE \textbf{Set:\:} $\mu>0$, $\beta>0$, $\mathbf{u}^{0}$, $\mathbf{v}^{0}$, $\mathbf{\lambda}^{0}$, $\epsilon$
		\STATE \textbf{While\:\:}stopping criterion\:$>\epsilon$\:\:\textbf{do}
		\STATE $\mathbf{u}^{i+1}=\operatorname{argmin}_{\mathbf{u}}\frac{1}{2} \parallel \mathbf{y} - \Phi \mathbf{u} \parallel_2^2+\frac{\beta }{2}\parallel \mathbf{u}-\mathbf{v}^i+\frac{\mathbf{\lambda}^i}{\beta} \parallel_2^2$
		\STATE $\mathbf{v}^{i+1}=\operatorname{argmin}_{\mathbf{v}}\frac{\mu}{2}\mathbf{v}^T(\mathbf{v}-\mathcal{F}(\mathbf{v}))+\frac{\beta }{2}\parallel \mathbf{u}^{i+1}-\mathbf{v}+\frac{\mathbf{\lambda}^i}{\beta} \parallel_2^2$
		\STATE $\mathbf{\lambda}^{i+1}= \mathbf{\lambda}^i+\beta(\mathbf{u}^{i+1}-\mathbf{v}^{i+1})$
		\STATE \textbf{End}
	\end{algorithmic}
\end{algorithm}
The denoiser used in the PnP algorithm does not relate to an explicit regularizer, and as such, the loss function is not explicitly defined. Romano~\textit{et al.} recently proposed another way, called RED, to exploit denoising algorithms for regularization~\cite{16M1102884} in which an explicit regularizer is defined as follows:
\begin{equation} 
\label{eq:12}
\rho(\mathbf{x}) = \frac{1}{2}\mathbf{x}^T(\mathbf{x}-\mathcal{F}(\mathbf{x})),
\end{equation}
where $\rho$ is designed to enforce the orthogonality of the image ($\mathbf{x}$) and what a denoiser removes from the image ($\mathbf{x}-\mathcal{F}(\mathbf{x})$). The main advantage of having an explicit regularizer in RED compared to PnP is that its theoretical convergence properties can be analyzed and different optimization algorithms can be used to solve the problem~\cite{8528509}. Furthermore, Romano~\textit{et al.} has shown numerical evidence that denoising algorithms are locally homogeneous under certain conditions which helps to compute the gradient expression of regularizer $\rho_{red}(.)$ as follows~\cite{16M1102884}:
\begin{equation} 
\label{eq:13}
\nabla \rho(\mathbf{x}) = \mathbf{x}-\mathcal{F}(\mathbf{x}),
\end{equation}
Herein, the proposed ADMM framework for ultrasound beamforming is extended to the RED algorithm. To do so, $\ell_1$-norm in Eq.~(\ref{eq:4}) is substituted by $\rho$ in Eq.~(\ref{eq:12}), and the resulting optimization problem is as follows:
\begin{equation} 
\label{eq:14}
\hat{\mathbf{x}} = \operatorname*{argmin}_{\mathbf{x}} \frac{1}{2}\parallel \mathbf{y} - \Phi \mathbf{x} \parallel_2^2+\frac{\mu}{2}\mathbf{x}^T(\mathbf{x}-\mathcal{F}(\mathbf{x})),
\end{equation} 
where its iterative solution is summarized in Algorithm~\ref{alg:3}. The only difference compared to original ADMM is the fifth line involving the new regularization term. Since $\rho$ is differentiable, the solution $\mathbf{v}^{i+1}$ must obey the fixed point relationship:
\begin{multline}
\label{eq:15}
\mu \nabla \rho_{red}(\mathbf{v})-\beta(\mathbf{u}^{i+1}-\mathbf{v}+\frac{\mathbf{\lambda}^i}{\beta})
=\\ \mu (\mathbf{v}-\mathcal{F}(\mathbf{v}))-\beta(\mathbf{u}^{i+1}-\mathbf{v}+\frac{\mathbf{\lambda}^i}{\beta})=0\;\;\;\;\;\;\;\;\;\;\;\;\;\;\;\;\;\;\;\;\;\;\;\\
\Leftrightarrow \mathbf{v^{i+1}} = \frac{\mu}{\mu+\beta}\mathcal{F}(\mathbf{v^{i+1}})+\frac{\beta}{\mu+\beta}\mathbf{u^{i+1}}+\frac{1}{\mu+\beta}\mathbf{\lambda^{i}},
\end{multline}
where the solution $\mathbf{v}^{i+1}$ is a function of its denoised version $\mathcal{F}(\mathbf{v^{i+1}})$. Therefore, an approximation of $\mathbf{v}^{i+1}$ can be obtained by iterating:
\begin{equation} 
\label{eq:16}
\mathbf{z^{k}} = \frac{\mu}{\mu+\beta}\mathcal{F}(\mathbf{z^{k-1}})+\frac{\beta}{\mu+\beta}\mathbf{u^{i+1}}+\frac{1}{\mu+\beta}\mathbf{\lambda^{i}}
\end{equation} 
over $k=1,...,K$ iterations with sufficiently large $K$. The previous value is used to initialize $\mathbf{z}$ (i.e., $\mathbf{z^0}=\mathbf{v^i}$). Similar to the PnP approach, NLM denoiser is used in the RED algorithm proposed in this work for ultrasound imaging. 
\section{Experiments}
\label{sec:sec3}
The experimental part is designed to provide a clear understanding of the advantages as well as the limitations of the proposed method. In this section, first, details regarding datasets on which the proposed method is evaluated and compared with other approaches are explained. Then, different criteria used for quantitative evaluation of results are introduced.
\subsection{Datasets}
\label{sec:sec31}
The performance evaluation is performed on publicly available PICMUS\footnote{The datasets are publicly available at PICMUS website:~\url{https://www.creatis.insa-lyon.fr/Challenge/IEEE IUS 2016/}} benchmark datasets~\cite{7728908}. It includes simulations, real phantoms, and in vivo datasets that are specifically designed for the assessment of ultrasound beamforming algorithms. This helps to have a fair comparison with other approaches and facilitates the future investigations. In short, each image of the dataset is reported as following:
\begin{enumerate}
	\item simulation resolution (SR): the image contains point targets distributed horizontally and
	vertically against an anechoic background designed to assess the spatial resolution.
	\item simulation contrast (SC): the image contains anechoic cysts distributed horizontally and
	vertically against fully developed speckle background designed to assess the contrast.
	\item experimental resolution (ER): the image was collected from a CIRS Phantom (Model 040GSE) corresponding to the regions containing wire targets over speckle background to assess the spatial resolution.
	\item experimental contrast (EC): this image was also collected from the same phantom as ER but in the regions containing anechoic cysts over speckle background to assess the contrast.
\end{enumerate}
The \textit{in vivo} images were collected from the carotid artery of a volunteer which include cross-sectional (denoted by Carotid Cross (CC)) and longitudinal views (denoted by Carotid Longitudinal(CL)). More details regarding simulation and imaging settings of PICMUS datasets can be found in~\cite{7728908}.\par
\subsection{Evaluation metrics}
\label{sec:sec32}
The reconstructed images using different beamforming methods are evaluated and compared in terms of resolution and contrast which are the main specialized ultrasound assessment indexes. Moreover, the performance of our method in preserving the speckle quality is also investigated.\par
Full Width at Half Maximum (FWHM), calculated in the axial as well as lateral directions, is used for resolution assessment. As for the contrast index, Contrast-to-Noise Ratio (CNR) is defined as follows:
\begin{equation} 
\label{eq:17}
CNR=20log_{10}(\frac{ \mid\mu_{ROI}-\mu_{B}\mid}{ \sqrt{(\sigma_{ROI}^2+\sigma_{B}^2)/2}}),
\end{equation} 
where $\sigma$ and $\mu$ denote the standard deviation and the mean parameters, respectively. Subscribes $._{ROI}$ and $._{B}$ refer to the region of interest and the background, respectively.\par 
Besides CNR, generalized CNR (gCNR)~\cite{8918059} is another contrast metric which has been shown to be robust against dynamic range alterations, and is calculated as following:
\begin{equation} 
\label{eq:18}
gCNR = 1- \int_{-\infty}^{\infty} min\left\{p_{ROI}(x),p_B(x)\right\}dx
\end{equation} 
where $p$ refers to the histograms of pixels in each region. gCNR quantifies the overlap between the distributions of pixels' intensities in two regions regardless of dynamic range transformations. Lower distributions overlap leads to higher gCNR values, and it is equal to the maximum value of 1 if the two distributions are disjoint~\cite{8918059,9662310}.\par
In ultrasound imaging, speckle is an important feature, and as such, it is crucial to preserve it during reconstruction. For the fully developed speckle regions, the intensity of resulting envelope image follows a Rayleigh distribution. Herein, a Kolmogorov–Smirnov (KS) test, designed by PICMUS organizers, is used to verify whether the data follows a Rayleigh distribution or not. In a region of image, the significance level of $\alpha = 0.05$ is considered to decide whether the speckle statistics are preserved or not.
\section{Results}
\label{sec:sec4}
This section is dedicated to the performance evaluation of proposed ADMM, PnP, and RED approaches. The images, introduced in Section~\ref{sec:sec31}, are reconstructed by solving the inverse problem of beamforming using the algorithms~\ref{alg:1},~\ref{alg:2}, and~\ref{alg:3}. Furthermore, the advantages of our methods are demonstrated in comparison with other approaches for solving the problem of ultrasound beamforming. Among the time-domain approaches, the classical DAS, EMV~\cite{5611687}, and phased coherence factor (PCF)~\cite{4976281} are selected for the sake of comparison. The results of Fourier domain technique based on Stolt’s migration~\cite{8359331}, and ultrasound Fourier slice beamforming (UFSB)~\cite{7582552} are also included. The comparison is completed by incorporating the results of our deep learning method, referred to as MNV2~\cite{9251565}, which was ranked first, in terms of image quality, in CUBDL challenge 2020~\cite{9475029}. Moreover, the beamforming methods are also compared in terms of the reconstruction time. Finally, the sensitivity of proposed methods to parameter selection and algorithm initialization is also analyzed.\par
Hereafter, the Hanning apodization window with f-number equals to 0.5 is always considered as reception apodization matrix in our method, DAS, and other methods on top of DAS (except for \textit{in vivo} datasets for which the Tukey (tapered cosine) window with constant parameter of 0.25 and f-number equals to 1.75 is considered). The search and comparison window sizes of NLM denoiser are set to 21 and 5, respectively. The reliable built-in MATLAB implementation of NLM is used in which the standard deviation of noise estimated from the image is used as degree of smoothing. This helps to have adaptive denoising with less parameters. Details regarding the method used for the estimation of noise variance can be found in~\cite{IMMERKAER1996300}. The threshold for stopping criterion is set to $\epsilon = 10^{-3}$. The initial values equal to zero are selected for all iterative algorithms. The hyperparameters of each method are tuned independently to get the best results. Quantitative indexes are separately calculated for each cyst region or point target in the image, and the average values are reported. 
\subsection{Beamforming results with the proposed approaches}
\label{sec:sec41} 
\begin{figure*}[t]
	\captionsetup{justification=centering}
	\centerline{\includegraphics[width=\textwidth]{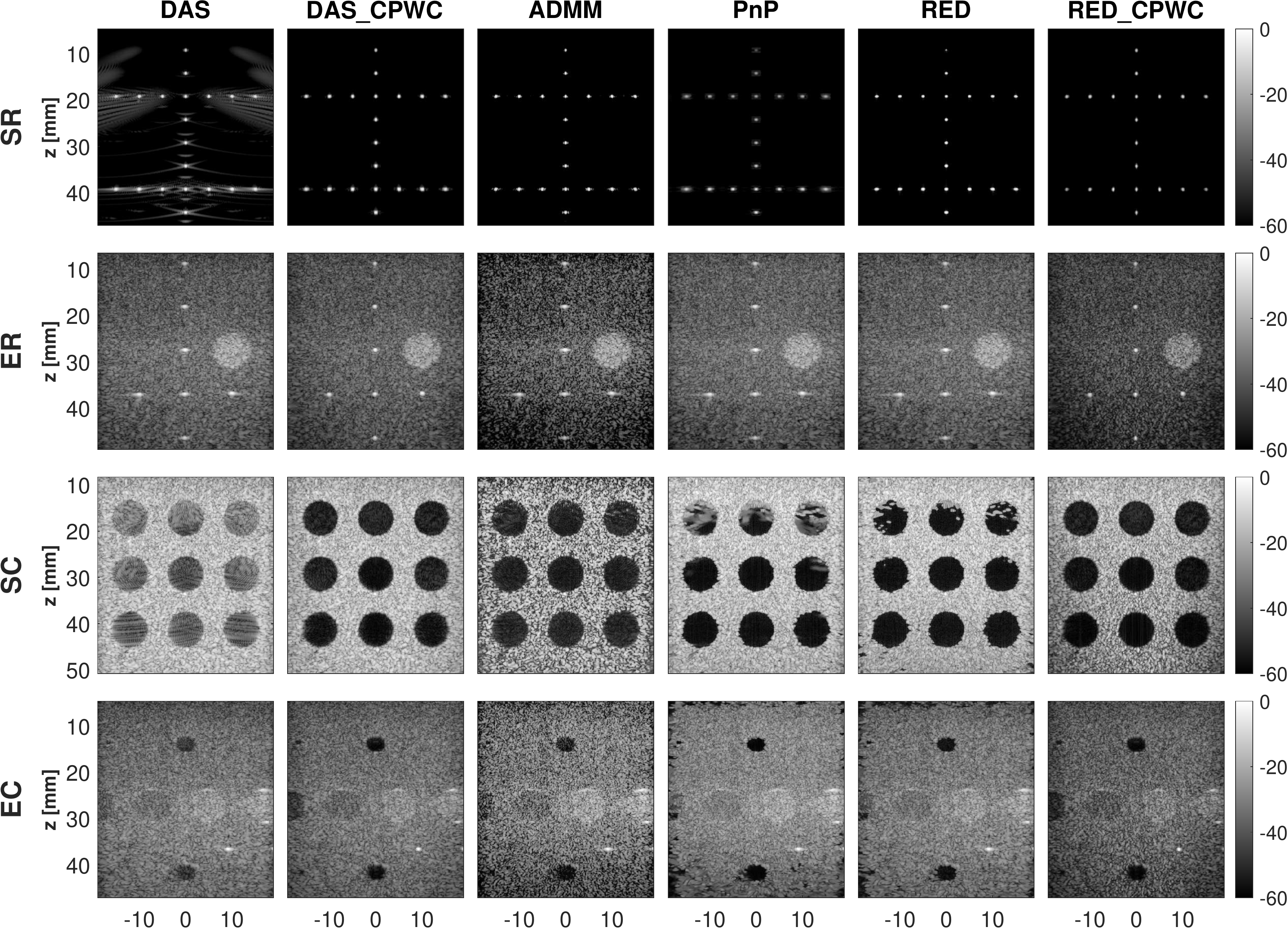}}
	\caption{The results of simulation and experimental datasets. Rows indicate datasets while columns correspond to different approaches. All results are from a single $0^o$ plane-wave insonification except for CPWC which is obtained from 75 steered insonifications.}
	\label{fig:fig1}
\end{figure*}
\begin{figure*}[t]
	\captionsetup{justification=centering}
	\centerline{\includegraphics[width=\textwidth]{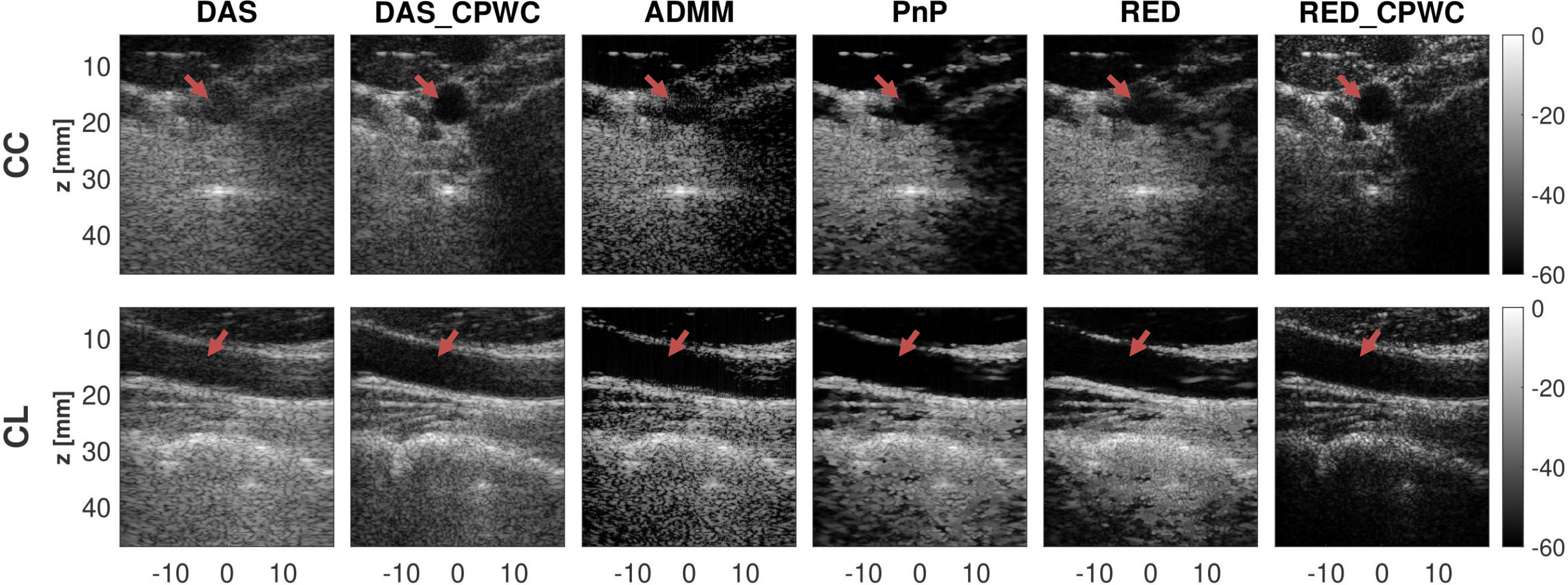}}
	\caption{The results of \textit{in vivo} datasets. Rows indicate datasets while columns correspond to different approaches. All results are from a single $0^o$ plane-wave insonification except for CPWC which is obtained from 75 steered insonifications.}
	\label{fig:fig2}
\end{figure*}
\begin{table}[t!]
	\caption{Quantitative results in terms of resolution and contrast indexes for simulation and real phantom experiments. The KS columns indicate whether the method preserves speckle texture or not, which are indicated by \cmark and \xmark ~marks, respectively.}
	\label{table:1}
	\centering
	\setlength{\tabcolsep}{1.5pt}
	\scriptsize
	\begin{tabular}{c c c c c c c c c c c c c c}
		\specialrule{.15em}{0em}{.2em}
		dataset & SR & ER & SC & EC  \\ [.2em] 
		\specialrule{.05em}{0em}{.2em} 
		index & FWHM\textsubscript{A} FWHM\textsubscript{L} & FWHM\textsubscript{A} FWHM\textsubscript{L} KS & CNR gCNR KS & CNR gCNR KS \\ [.2em] 
		\specialrule{.05em}{0em}{.2em} 
		\makecell{DAS \\ DAS\_CPWC \\ ADMM \\ PnP \\ RED \\ RED\_CPWC} & \makecell{0.4 \\ 0.4 \\ 0.38 \\ 0.29 \\ 0.37 \\ 0.36} \makecell{0.47 \\ 0.4 \\ 0.39 \\ 0.43 \\ 0.46 \\ 0.26} & \makecell{0.48 \\ 0.49 \\ 0.47 \\ 0.48 \\ 0.48 \\ 0.46} \makecell{0.8 \\ 0.55 \\ 0.74 \\ 0.75 \\ 0.76 \\ 0.36}  \makecell{\cmark \\ \cmark \\ \xmark \\ \cmark \\ \cmark \\ \cmark} & \makecell{10.25 \\ 17.53 \\ 9.01 \\ 14.87 \\ 15.48 \\ 18} \makecell{0.89 \\ 0.99 \\ 0.88 \\ 0.93 \\ 0.94 \\ 1} \makecell{\cmark \\ \cmark \\ \xmark \\ \cmark \\ \cmark \\ \cmark} &  \makecell{8.8 \\ 13.25 \\ 6.85 \\ 16.45 \\ 14.7 \\ 15} \makecell{0.87 \\ 0.97 \\ 0.72 \\ 0.99 \\ 0.98 \\ 1} \makecell{\cmark \\ \cmark \\ \xmark \\ \cmark \\ \cmark \\ \cmark} \\ [.2em] 
		\specialrule{.05em}{0em}{.2em}
		\makecell{EMV \\ PCF \\ MNV2 \\ Stolt \\ UFSB} & \makecell{0.4 \\ 0.3 \\ 0.42 \\ 0.42 \\ 0.4} \makecell{0.1 \\ 0.38 \\ 0.27 \\ 1.1 \\ 0.85} &    \makecell{0.59 \\ 0.46 \\ 0.53 \\ 0.55 \\ 0.55} \makecell{0.42 \\ 0.41 \\ 0.77 \\ 0.41 \\ 0.52} \makecell{\cmark \\ \xmark \\ \cmark \\ \cmark \\ \cmark }& \makecell{11.21 \\ 5.64 \\ 10.48 \\ 2.1 \\ 7.3} \makecell{0.93 \\ 0.76 \\ 0.89 \\ 0.55 \\ 0.78} \makecell{\cmark \\ \xmark \\ \cmark \\ \xmark \\ \xmark }&  \makecell{8.1 \\ 3.2 \\ 7.8 \\ 6.55 \\ 5.96} \makecell{0.83 \\ 0.68 \\ 0.83 \\ 0.78 \\ 0.76} \makecell{\cmark \\ \xmark \\ \cmark \\ \cmark \\ \cmark }\\ [.2em] 
		\specialrule{.05em}{0em}{.2em}
	\end{tabular}
\end{table}
\subsubsection{Simulation and experimental data}
\label{sec:sec411} 
\begin{figure*}[t]
	\captionsetup{justification=centering}
	\centerline{\includegraphics[width=0.78\textwidth]{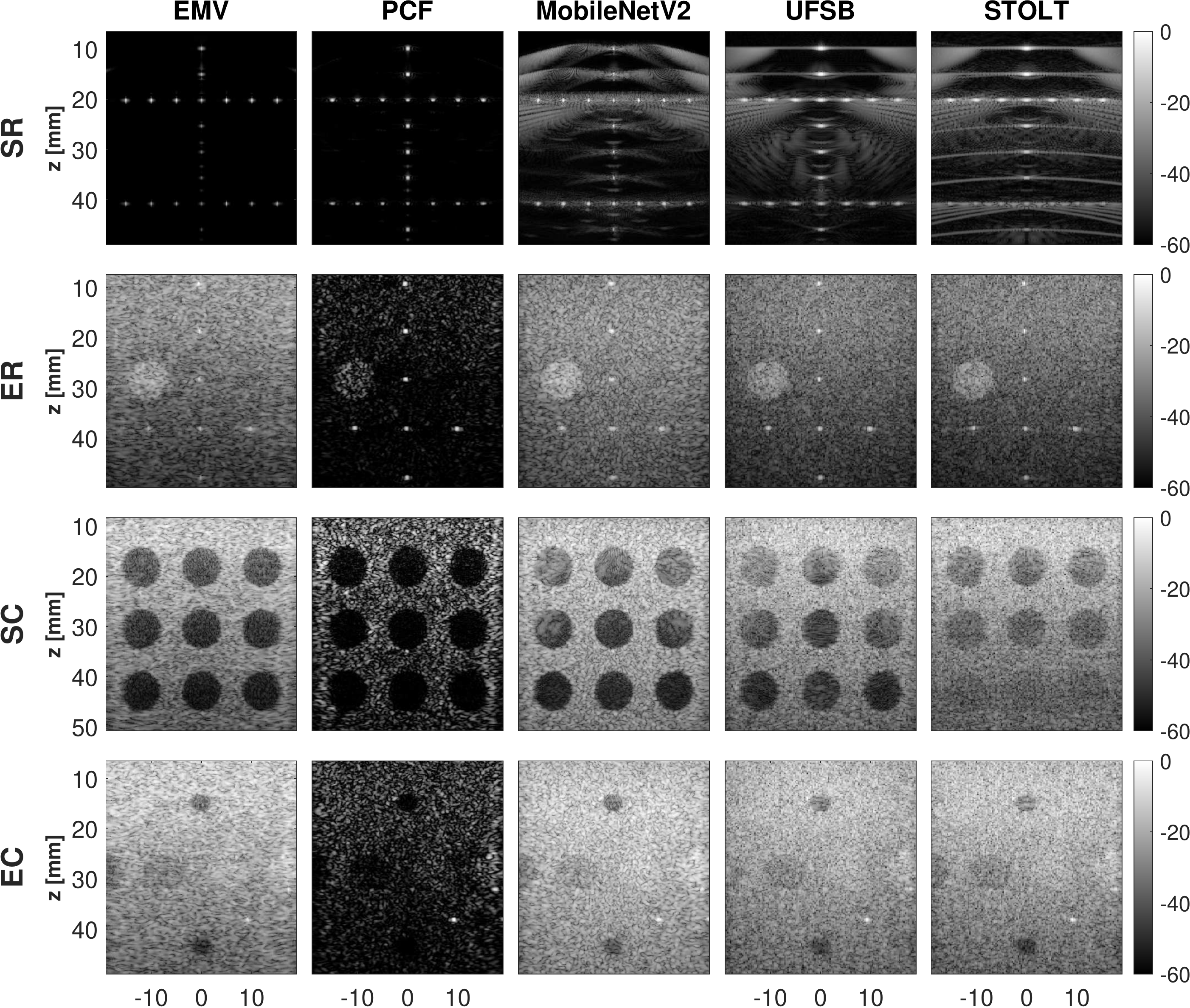}}
	\caption{Simulation and experimental images reconstructed through previous beamforming methods. Rows indicate datasets while columns correspond to different approaches. All results are from a single $0^o$ plane-wave insonification.}
	\label{fig:fig3}
\end{figure*}
Fig.~\ref{fig:fig1} displays the beamforming results from the single $0^o$ plane-wave as well as Coherent Plane-Wave Compounding (CPWC) on 75 insonifications. The quantitative evaluation is also summarized in Table~\ref{table:1}. As results confirm, the PnP and RED approaches substantially improve the contrast and perfectly suppresses the side-lobe artifacts similar to DAS\_CPWC which averages the DAS results over 75 angles. The ADMM approach fails the KS test due to the soft-thresholding operator while PnP as well as RED algorithms replace it with denoising and preserve the speckle statistics.\par 
To apply the proposed approaches on different insonification angles, the process is as follows. First, the matrix $\Phi$ is independently calculated for each of the transmitting angles, given that the traveling time of the tilted plane-wave to the scatterers is different. Then, the proposed algorithms are independently performed for each angle. Finally, the resulting RF data (before envelope detection) are averaged to coherently compound all plane waves. Since the procedure is exactly the same for ADMM, PnP, and RED algorithms, we only present the results of RED algorithm, on 75 insonification angles, in order to keep the paper concise. The RED\_CPWC results confirm that the method is extensible to different angles, and CPWC improves the results as compared to a single $0^o$ insonification.\par
The proposed methods marginally improve the resolution as compared to DAS. This result is expected because the focus of ADMM, PnP, and RED algorithms is on the regularization term which mainly affects the contrasts of results. The resolution, however, is rooted in the measurement model and mainly depends on the matrix $\Phi$. Nevertheless, the visual inspection of SR image reveals that PnP method makes the point targets blurry due to the averaging of image patches in NLM denoiser. RED algorithm does not suffer from this problem since the denoiser is used to define an explicit regularizer rather than directly applying on the image.
\subsubsection{\textit{In vivo} data}
\label{sec:sec412}
Fig.~\ref{fig:fig2} illustrates the cross-sectional and longitudinal views from the carotid artery. The visual comparison of different approaches confirm that the clutter artifacts caused by diffuse reverberation from shallow layers are suppressed by proposed algorithms, and dark images of the artery are reconstructed in both views (pointed out by red arrows).
\subsection{Comparison with existing beamformers}
\label{sec:sec42}
Fig.~\ref{fig:fig3} shows the results of existing state of the art beamforming methods with the quantitative results reported in Table~\ref{table:1}. Overall, it can be verified that EMV beamformer is of the highest axial resolution among all approaches including ours. But the proposed RED algorithm is the best in terms of contrast. Another advantage of RED is that similar contrast indexes are achieved for both simulation and experimental phantoms. However, EMV performance noticeably drops in real experiments as compared to simulation. This difference is mainly brought about by the selection of eigenvectors of the covariance matrix used for making the signal subspace. More specifically, the SR data includes anechoic background from which the point targets can be recovered by only using the principal eigenvectors. However, the reconstruction of ER image requires considering all eigenvector in order not to lose the speckle texture, which makes the results of EMV exactly the same as MV. This difference is also visible for contrast because experimental data contains an additional noise reducing the quality of estimated covariance matrix.\par
In terms of computational time, Fourier domain beamforming techniques are the fastest options, and the iterative approaches are the slowest ones. Herein, we compare the reconstruction time of different algorithms for EC experiment. The Stolt’s migration and UFSB approaches take 100 milliseconds and 2.92 seconds, respectively. Among the time-domain approaches, DAS and PCF have a similar speed and take 1.6 seconds to reconstruct the image. The EMV method is much slower as it needs 8 minutes to form the image mainly due to the covariance matrix estimation and its decomposition. MNV2 is the optimized version of MV which reduces the time to 40.2 seconds. The proposed ADMM, PnP, and RED approaches are even slower than EMV mainly due to the numerical optimization method used for optimizing the measurement loss and avoiding intractable matrix inversion. Although all of the proposed approaches converge in less than 15 iterations, the required time is around 20 minutes because each iteration contains an inner iterative algorithm (BFGS). It has to be mentioned that all these numbers are with straightforward Matlab implementations without any runtime optimization.

\subsection{Sensitivity analysis}
\label{sec:sec44} 
As mentioned in Algorithms~\ref{alg:1},~\ref{alg:2}, and ~\ref{alg:3}, initial values must be set for new variables (i.e., $\mathbf{u}$ and $\mathbf{v}$) and the Lagrange multiplier ($\lambda$). Moreover, the hyperparameters $\mu$ and $\beta$, which respectively determine the coefficient of regularizer and penalty term, should be specified.\par
Fortunately, whatever initialization for new variables and the Lagrange multiplier do not change the final solution since Eq.~(\ref{eq:4}) is convex and does not have any local minimum. But the number of iterations required for convergence might alter if outlying values are selected. Herein, the initial points are always set to zero.\par
As for the hyperparameters, a large $\beta$ (equals to 1000) is set in order to perfectly accomplish the equality constraints of new variables (Eq.~(\ref{eq:5})) and converge toward equal values for $\mathbf{u}$ and $\mathbf{v}$. While the regularizer coefficient $\mu$ is not used within PnP, it specifies the threshold of Eq.~(\ref{eq:10}) in ADMM algorithm. Large $\mu$ makes the ADMM results too dark since it wipes out the speckle texture while small $\mu$ cause a poor contrast. In RED approach, $\mu$ must be set to high values (between 1000 to 5000) as the denoiser residual ($\mathbf{x}-\mathcal{F}(\mathbf{x})$) is of a negligible amplitude. RED has also an extra parameter $K$ which specifies the number of iterations in Eq.~(\ref{eq:16}). Although sufficiently large $K$ is required for the exact RED algorithm, our investigations demonstrate that many inner iterations are unnecessary and $K=1$ is enough to get proper results. This behavior has also been reported in~\cite{8528509}.
\section{Discussions}
\label{sec:sec5}
The efficacy of inverse problem formulation of ultrasound beamforming has already been shown in several studies~\cite{7565515,8052532,8091286,7728907}. The main issue, however, is the loss of speckle texture in the proximal mapping step. The proposed RED algorithm is a reliable solution to this problem. The RED approach not only takes advantage of denoising for image reconstruction but also explicitly minimizes a well-defined regularizer. RED’s ability to preserve the speckle information is of crucial significance in image computing applications such as quantitative ultrasound and speckle tracking.\par
Once the medium is insonified, the spherical waves get backscattered toward the probe elements. Signals originating from pixels located in shallow regions are recorded by a small subset of elements close to the pixel location, while the backscattered signals originating from the subordinate pixels are properly recorded by most of the elements. That is why matrix $\Phi$ is multiplied with a reception apodization matrix to keep the f-number fixed for the entire image depths. Therefore, the anechoic cysts of shallow regions are reconstructed with a small part of probe elements which causes a lower contrast and brings some artifacts, as seen in the results of SC datasets. Nevertheless, those artifacts disappear in the result of CPWC where more data is available.\par
The modular property of ADMM helped us design a single framework for three different approaches of solving the inverse problem of ultrasound beamforming. Besides changing the prior term, ADMM provided the possibility of optimizing the measurement loss using limited-memory BFGS in which the Hessian matrix is approximated. This property is essential in our case wherein the calculation of Hessian is intractable due to the large size of matrix $\Phi$.\par
Although we have only presented the results for NLM denoiser, another important advantage of the proposed PnP and RED approaches is to open a way to incorporate any denoiser algorithm in the inverse problem of beamforming which is our plan for future work. This point is worth noting because the range of denoiser options is much larger than the choice of regularization functions.\par
Although using the noise standard deviation, estimated from the image, as degree of smoothing helps to have adaptive denoising with less parameters, it may cause loss of structural details in the low quality regions of the image. As seen in top regions of the in vivo datasets as well as top and left edges of the EC dataset, the image quality is low and the noise standard deviation is overestimated in those regions, which results in stronger smoothing and removes some structural details. This point is unavoidable unless the degree of smoothing is controlled manually, or a better noise variance estimation algorithm is adopted. The latter represents an interesting avenue for future work.\par
Achieving the same contrast as CPWC can be considered as a step toward eliminating the necessity of transmitting several plane-waves with different angles. However, the improvement in resolution is minor and we still need to extend the measurement model and the way matrix $\Phi$ is defined, which is the subject of our future research.\par 
Our formulation can be applied to other imaging techniques (such as focused and synthetic aperture imaging) or even other types of ultrasound transducer (such as convex and phased array). Herein, we only consider reporting the results on benchmark PICMUS dataset since it is available online which makes the comparison with previous approaches easier. It also facilitates reimplementing the methods and verifying the results. 
\section{Conclusions}
\label{sec:sec7}
Denoising algorithms have been recently adopted in solving the inverse problem of imaging. Herein, we proposed a novel framework for incorporating denoisers in medical ultrasound beamforming. Our framework is based on the ADMM wherein a linear forward model is used for the image under scrutiny, and three solutions are found by considering Laplacian, PnP, and RED priors. The results show that the proposed RED approach gives the best images quality with a high contrast while the speckle information is also preserved. 
\section{Acknowledgment}
\label{sec:sec8}
We thank the organizers of the ultrasound toolbox and the PICMUS challenge for providing publicly available data and codes. The authors sincerely thank Mohammed Albulayli and Adrien Besson for making their MATLAB codes available online.
\bibliographystyle{IEEEbib}
\bibliography{refs}
\end{document}